\documentstyle[aps, prl, epsfig, twocolumn]{revtex}

\begin{document}

\draft

\title{Winding up by a quench: vortices in the wake of rapid Bose-Einstein 
condensation}
\author{J.R. Anglin$^{1,2}$\thanks{e-mail: james.anglin@uibk.ac.at} and 
W.H. Zurek$^2$\thanks{e-mail: whz@lanl.gov}}
\address{$^1$Institut f\"ur Theoretischephysik, Universit\"at Innsbruck, 
Technikerstrasse 25, 6020 Innsbruck, Austria\\
$^2$T-6 (Theoretical Astrophysics), MS B288, Los Alamos National Laboratory, 
Los Alamos, New Mexico 87545}

\maketitle

\begin{abstract} A second order phase transition induced by a rapid quench can
lock out topological defects with densities far exceeding their equilibrium
expectation values.  We use quantum kinetic theory to show that this mechanism,
originally postulated in the cosmological context, and analysed so far only on
the mean field classical level, should allow spontaneous generation of vortex
lines in trapped Bose-Einstein condensates of simple topology, or of winding
number in toroidal condensates.  
\end{abstract}


\narrowtext

An as yet unachieved goal of experiments on trapped ultra-cold alkali
gases\cite{BECexp} is the exhibition of a persistent vortex.  Since the reason
that superfluid vortices are persistent is that there is a high energetic
barrier between the metastable vortex state and the non-rotating true ground
state, spinning up a non-rotating condensate once it is fully grown seems likely
to be difficult to accomplish without excessive heating.  In this Letter we show
that a rotating condensate may instead grow spontaneously from fluctuations
during a non-equilibrium quench.  Not only is this a possible procedure
for generating vortices:  it actually provides a strong additional motivation
for vortex experiments, by making vortex production in cooled alkali gases a
test of a fundamental prediction of non-equilibrium statistical mechanics.

The widely applied time-dependent Ginzburg-Landau
theory (TDGL) predicts failure of equilibrium during a second order phase
transition at finite speed.  In TDGL, the complex order
parameter $\psi(\vec{r},t)$ obeys
\begin{equation}\label{TDGL} 
\tau_0\dot\psi = \beta\Bigl({\hbar^2\over2M}\nabla^2 +\mu -\Lambda 
|\psi|^2\Bigr)\psi\;, 
\end{equation} 
where $\beta=(k_B T)^{-1}$, and $\tau_0$ and $\Lambda>0$ are 
phenomenological parameters.  The thermodynamical variable $\mu$ behaves 
near the critical point, in the case we consider, as
\begin{equation}\label{muTc}
\mu = {3\over2}(T_c -T) + {\cal O}(T_c-T)^2\;,
\end{equation}
where $T_c$ is the critical temperature.  The equilibration time for long 
wavelengths is $\tau=\tau_0 k_BT/|\mu|$.  The system's
disordered phase is described by $\mu<0$, so that $\psi=0$ is a stable fixed 
point of (\ref{TDGL}).
The ordered phase appears when $\mu>0$, since then the stable fixed points lie 
on the circle
$|\psi|^2=\mu/\Lambda$, and the phase $\theta$ of $\psi=|\psi|e^{i\theta}$ 
becomes a new macroscopic variable.

A quench occurs if $\mu$ changes with time from negative to positive values.
The divergence of the equilibration time $\tau$ at the critical point $\mu=0$ is
associated with {\it critical slowing down}.  Because of this critical slowing
down, ${d\mu\over dt}/\mu$ must exceed $1/\tau$ in some neighbourhood of the 
critical
point, and so there must be an epoch in which the system is out of equilibrium.
What are at the beginning of this epoch mere fluctuations in the disordered
phase, in which higher energy modes happen momentarily to be more populated than
the lowest mode, can thus pass unsuppressed by equilibration into the ordered
phase, to become topologically non-trivial configurations of the order parameter
field $\psi$.  One therefore expects topological defects, such as vortex lines,
to form spontaneously during a transition at sufficient speed\cite{WHZ}.

The interval within which equilibration is negligible can be identified as the
period wherein $|t|/\tau <1$.  If we define the quench time scale $\tau_Q$ by
letting $\beta\mu = t/\tau_Q$ (choosing $t=0$ as the moment the system crosses
the critical point), this implies that the crucial interval is
$-\hat{t}<t<\hat{t}$, for $\hat{t} = \sqrt{\tau_Q\tau_0}$\cite{WHZ}.  The
correlation length $\hat{\xi}$ for fluctuations at time $t=-\hat{t}$ is then
given by $\hbar/(2M\hat{\xi}^2)=\mu(-\hat{t})$, which (assuming
$T(-\hat{t})\doteq T_c$) implies that $\hat{\xi}=\lambda_{T_c}
(\tau_Q/\tau_0)^{1/4}$, for $\lambda_{T}=\hbar(2Mk_BT)^{-{1\over2}}$ the thermal
de Broglie wavelength.  Taking this correlation length as giving the typical
domain size surrounding a defect\cite{Kibble} implies that the vortex line
density, in bulk, should be proportional to $\tau_Q^{-1/2}$\cite{WHZ}.
Alternatively one can consider the transition to occur within a toroidal vessel,
so that independent random settings of the order parameter phase, at different
points around the torus, can produce a net {\it vorticity},
$W={1\over2\pi}\oint\!dl\,\nabla \theta$.  This implies a superflow, with
velocity $\hbar\vec{\nabla}\theta/M$\cite{footnote}.  In this case one estimates
one independently chosen phase within each correlation length $\hat \xi$;
modeling the phase distribution around the torus as a random walk suggests that
the net vorticity should be proportional to $\hat{\xi}^{-1/2}$, hence to
$\tau_Q^{-1/8}$\cite{WHZ}.

Although ingenious experiments have recently been performed to test this theory,
in liquid helium\cite{HE3exp}, and numerical studies have supported its scaling
predictions\cite{numerical}, it would be even more informative to have analogous
results in a weakly interacting system, such as a dilute trapped alkali gas.
Assuming $\tau_0$ is the scattering time, evaporative cooling techniques yield
$(\tau_Q/\tau_0)^{1/4}$ of order one, and so $\hat{\xi}$ is essentially
$\lambda_{T_c}$.  For atoms at several hundred nK, this means $\hat{\xi}\sim
100$ nm, smaller than current condensates.  As numerical simulations
show\cite{numerical}, this is a generously low lower bound on the distance
between vortex lines, but it does indicate that spontaneous vorticity should be
within experimental reach.  Considering this intriguing prospect raises an
obvious question:  is TDGL actually relevant to finite samples of dilute gas,
far from equilibrium?

We therefore begin again from first principles,
and consider a dilute Bose gas in a trap, with the
Hamiltonian 
\begin{eqnarray}\label{H} 
\hat{H} = {\hbar^2\over2 M}
\int\!d^3r\,\Bigl(|\vec\nabla\hat\psi|^2 
	+ U(\vec{r})\hat\psi^\dagger\hat\psi +
	4\pi a \hat\psi^{\dagger 2}\hat\psi^2\Bigr)\; 
\end{eqnarray}
where $\hat\psi(\vec{r})$ annihilates a boson at position $\vec r$, $U$ gives
the trap potential, and $a$ is the s-wave scattering length.  As always,
$\hat{\psi}(\vec{r})=\sum_ku_k(\vec{r})\hat{\psi}_k$ defines a decomposition of
the system into orthogonal modes described by single-particle wave functions
$u_k$.  In the earliest stages of condensation, it is sufficient to take the
single-particle energy eigenstates as defining the normal modes of the gas.

We now construct a quantum kinetic theory (QKT), by considering the lowest
energy modes of the trap, up to some energy $E_R$, to be an open
quantum system (the `condensate band'), interacting via two-particle
s-wave scattering with the higher modes, treated as a `reservoir 
band'\cite{QKT}.  We
model evaporative cooling by prescribing that the reservoir band is always
in equilibrium, but with a time dependent temperature
$\beta^{-1}(t)$ and chemical potential $\mu(t)$ (which can become positive as 
long as it remains below $E_R$).  We then form the
reduced density operator for the condensate band by tracing out the
reservoir.  The condensate band will not remain in equilibrium with the
reservoir; the time evolution of its reduced density operator is
the problem to be solved.

In the earliest stages of condensation, before nonlinear coherent
interactions become important, one can
derive a simple master equation for the condensate band,
strongly reminiscent of that of a multi-mode laser:
\begin{eqnarray}\label{ME1}
\dot{\hat\rho} &=& \sum_k \Bigl({E_k\over i\hbar} [\hat{n}_k, \hat\rho] + 
\Gamma_k e^{\beta\mu}\Bigl[e^{\beta(E_k-\mu)} \hat{a}_k\rho\hat{a}_k^\dagger +	
	\hat{a}_k^\dagger\rho\hat{a}_k \nonumber\\
&&\ \ \ \  -{1+e^{\beta(E_k-\mu)}\over2}(\hat{n}_k\hat{\rho} + 
\hat{\rho}\hat{n}_k) - 
\hat{\rho}\Bigr]\Bigr)\;, 
\end{eqnarray}
where $E_k$ are the energies of the normal modes.  The $\Gamma_k$ are
scattering rates, which may be computed;
they will generally be of the order of the Boltzmann scattering rate.  We 
actually expect the $k$-dependence of the $\Gamma_k$ to be
weak as long as the temperature is much larger than the trap level
spacing, so we will hereafter replace $\Gamma_k$ with
$\Gamma_0$, which will play exactly the same role as $1/\tau_0$ did in TDGL.  
The non-Hermitian part of (\ref{ME1}) is
due to collisions in which one particle leaves or joins the condensate
for or from the reservoir.  

An ansatz which solves (\ref{ME1}) is furnished by
\begin{equation}\label{prod}
\hat\rho(t) = \prod_k {1\over \bar n_k + 1} \sum_{n_k} 
	\Bigl({\bar n_k\over\bar n_k +1}\Bigr)^{n_k} 
				|n_k\rangle\langle n_k|\;,
\end{equation}
where $\bar n_k(t) = \hbox{Tr}(\hat\rho \hat n_k)$.  The equation
governing the $\bar n_k(t)$ follows simply from (\ref{ME1}):
\begin{equation}\label{nev}
\dot{\bar n}_k = \Gamma_0 e^{\beta\mu}
		[1+ (1-e^{\beta(E_k-\mu)})\bar n_k]\;.
\end{equation}
This equation may be integrated for general $\beta(t),
\mu(t)$. But as a simple form valid near the critical point, we
impose $\beta(t)[\mu(t)-E_k] = (t-\vartheta_k)/\tau_Q$, defining $\tau_Q$ as
well as the bias time scales $\vartheta_k$.  The $\bar{n}_k(t)$ that result,
from the equilibrium initial values $\bar{n}_k(t_i) = (e^{\beta(t_i)[E_k
- \mu(t_i)]}-1)^{-1}$, are incomplete Gamma functions; they
only begin to depart significantly from their instantaneous equilibrium
values after $t - \vartheta_k \simeq -\sqrt{\tau_Q/\Gamma_0}=-\hat{t}$.  Past 
these
points, the $\bar{n}_k$ lag below their equilibrium values.  This clarifies
the effect of the critical slowing down: as Bose enhancement turns on, the rates 
of scattering into the condensate increase; but the
numbers of particles required by equilibrium increase faster still,
and so the ability of scattering to maintain equilibrium
rapidly declines.

After these times, we can approximate $(1-e^{\beta(E_k-\mu)})\doteq 
(t-\vartheta_k)/\tau_Q$ and match to equilibrium at early times, to see that
\begin{equation}\label{nk}
\bar{n}_k(t) \doteq \Gamma_0 e^{{1\over2\hat{t}^2} 
(t-\vartheta_k)^2}\int_{-\infty}^{t-\vartheta_k}\!dt'\,
	e^{-{1\over2\hat{t}^2}t^{'2}}\;.
\end{equation}
For times after $t-\vartheta_k\simeq \hat{t}$, each $\bar{n}_k$ grows 
explosively,  
because the atomic scattering analogue of
stimulated emission into the $k$th mode is turning on strongly:  $\bar{n}_k$ is
becoming large enough that the term proportional to it on the RHS of (\ref{nev})
dominates the other term.  Bose-enhanced scattering then enables the mode to
begin a very rapid `whiplash' to catch up with equilibrium.  So the interval
$\vartheta_k-\hat{t}<t<\vartheta_k+\hat{t}$ is indeed a transition zone 
between
equilibrium above $T_c$, and the onset of
coherent processes below $T_c$.  It is obvious that for a
higher energy mode to have any significant chance of competing successfully for
particles with the lowest mode, it cannot afford to begin explosive growth much
later than the lowest mode.  This implies that $\vartheta_k < \hat{t}$, or 
$\beta E_k
< (\Gamma_0\tau_Q)^{-1/2}$, limits the range of significantly competitive modes.
Since in bulk or in a toroidal trap we have $E_k\propto k^2$, this gives
\begin{equation}
\hat{\xi}={1\over\hat{k}} = \hbar(2Mk_BT_c)^{-{1\over2}}(\Gamma_0\tau_Q)^{1/4}
\end{equation}
which is the same conclusion reached by TDGL\cite{WHZ}.  

For the toroidal problem, the density operator prescribed by the linear quantum
kinetic theory is equivalent to a distribution of coherent states with
probabilities proportional to $\exp-\sum_k{1\over\bar{n}_k}|\psi_k|^2$, for
Fourier modes $k$.  While $W$ is {\it not} a simple function of $\psi_k$, the
idea that there are as many independent random phases as non-negligible
$\bar{n}_k(\hat{t})$ still seems reasonable, and we expect typical vorticities
of order $\sqrt{\hat{k}L/2\pi}$, for $L$ the perimeter of the torus.  This 
again co-incides with the TDGL prediction.

Our conclusion at this point is that QKT agrees with 
the phenomenological theory, in predicting that for sufficiently rapid
quenches the probability of forming a small `seed' of condensate with
non-zero vorticity is of order one.  But since superfluid currents only
become metastable above a threshold condensate density, not all of this
initial vorticity will survive as the condensate grows.  To follow the 
non-equilibrium evolution of a trapped condensate into the non-linear regime, 
with quantum kinetic theory, is a challenging problem.  We therefore 
restrict our analysis to a simple toy model, which affords some 
qualitative insight, and allows a comparison between TDGL
and QKT.

The toy model replaces the condensate band of many low energy modes by
a system with only two modes, representing states with two different
angular momenta.  Because the self-Hamiltonian for this two-mode system
must conserve both particle number and angular momentum, it must
conserve separately the numbers of particles in both modes.  We
therefore choose
\begin{equation}\label{twomode}
\hat{H} = E [\hat{n}_1 + {1\over 2N_c} (\hat{n}_0^2 + \hat{n}_1^2 + 4 
\hat{n}_1\hat{n}_0)]\;.
\end{equation}
Because we have incorporated the Bose enhancement of inter-mode
repulsion (the factor of 4 instead of 2 in front of the
$\hat{n}_1\hat{n}_2$ term, which is of course the best case value,
obtained when $u_0$ and $u_1$ overlap completely), we make the state
with all particles in the 1 mode a local minimum of the energy for $n_1
+ n_2 > (N_c+1)$.  For two lowest modes of a typical oblate magneto-optical
trap, we have $\beta E$ of order $10^{-2}$; for proposed toroidal traps with 
perimeter of order $10^{-2}$ cm, at similar temperatures, $\beta E$ could be as 
low as $10^{-5}$.  (Rotating the gas before condensation could also lower the 
effective energy bias, and even favour rotating states over the ground state.)  
The experimental range of $N_c$ is around 100 for compact 
traps, but as low as 1 for the torus; this does not take into account the 
Thomas-Fermi expansion of the condensate wave function, which in fact can make 
$N_c$ rise significantly at large particle numbers.

We also assume interactions between both condensate modes and the
quasi-continuum of reservoir modes, of the form implied by the
Hamiltonian (\ref{H}).  Upon tracing over the dilute gas reservoir,
we obtain a master equation of more complicated form than
(\ref{ME1}), which includes saturation effects, as well as scattering of 
reservoir atoms off the condensate (with no resulting change in the condensate 
number).  For present purposes only the diagonal part of this equation is 
necessary:
\begin{eqnarray}\label{ME2}
\dot{p}_{n_0,n_1} &=& - \Gamma(t)[R_{n_0,n_1}-R_{n_0-1,n_1} + S_{n_0,n_1} - 
S_{n_0,n_1-1}]\nonumber\\
	&&-\tilde{\Gamma}(t)[T_{n_0+1,n_1} - T_{n_0, n_1+1}]\nonumber\\
R_{n_0,n_1} &\equiv& (n_0+1)[e^{\beta\mu}p_{n_0,n_1}
	-e^{{\beta E\over N_c}(n_0+2n_1)}p_{n_0+1,n_1}]\nonumber\\
S_{n_0,n_1} &\equiv& (n_1+1)[e^{\beta\mu}p_{n_0,n_1}
	-e^{{\beta E\over N_c}(N_c+2n_0+n_1)}p_{n_0,n_1+1}]\nonumber\\
T_{n_0,n_1} &\equiv& n_0n_1 e^{-{1\over2}{\beta E\over N_c}|N_c+n_0-n_1|}
	[e^{{1\over2}{\beta E\over N_c}(N_c+n_0-n_1)}p_{n_0-1,n_1}\nonumber\\
&&\ \ -e^{-{1\over2}{\beta E\over N_c}(N_c+n_0-n_1)}p_{n_0,n_1-1}]\;,
\end{eqnarray}
where $\Gamma(t)$ and $\tilde\Gamma(t)$ are again scattering rates (for
scattering into/out of the condensate, and off the condensate, respectively)
which may be computed for any specific condensate-reservoir coupling.  We will
hereafter assume $\tilde\Gamma=\beta E\Gamma$, which is accurate for simple trap
configurations when the temperature is much larger than the trap level spacing.
(This $\beta E$ factor justified neglecting these bouncing-off processes in the
linear regime; it appears because most reservoir particles are so much faster
than the condensate particles that they are unlikely to strike them without
dislodging them from the condensate band.)

Equation (\ref{ME2}) provides a complete description of condensation in the toy
model, including initial seeding from fluctuations, coherent growth, relaxation
into metastable states, and eventual equilibration by thermal barrier crossing.
While it would be straightforward to solve numerically, we can obtain more
understanding of the growth process by extracting from it an equation of motion
for $n_0$ and $n_1$.  This may be done, among other ways, by taking $n_0\to Nx$
and $n_1\to Ny$ for continuous $x$ and $y$ and $N$ of order $(\beta E)^{-1}$.
Expanding the finite differences in (\ref{ME2}) in powers of derivatives with
respect to $x$ and $y$, one obtains a Fokker-Planck-like equation, the Liouville
terms of which describe a flow along deterministic trajectories in
$(x,y)$-space.  Dropping higher order terms in $1/N$ (since these are
significant only at small $n_0,n_1$, when diffusion dominates systematic
evolution but we are able to use the linear analysis described above), these
trajectories obey 
\begin{eqnarray}\label{traj} 
\dot{n}_0 &=& \Gamma  n_0
\Bigl[e^{\beta\mu} -e^{{\beta E\over N_c}(n_0+2n_1)}\nonumber\\ 
&&+2\beta E n_1 e^{-{\beta E\over2N_c}|N_c+n_0-n_1|} 
	\sinh{\beta E\over2N_c}(N_c+n_0-n_1)\Bigr]\nonumber\\ 
\dot{n}_1 &=& \Gamma n_1\Bigl[e^{\beta\mu} 
	-e^{{\beta E\over N_c}(N_c+n_1+2n_0)}\\ 
&&-2\beta E n_0 e^{-{\beta E\over2N_c}|N_c+n_0-n_1|} 
	\sinh{\beta E\over2N_c}(N_c+n_0-n_1)\Bigr]\nonumber .  
\end{eqnarray}

The first question is, how important is the systematic evolution prescribed by
(\ref{traj}) compared to the diffusive evolution also contained in (\ref{ME2})?
We can address this question by examining a Gaussian approximation to the
Fokker-Planck equation from which (\ref{traj}) came.  Fig.~1 shows selected
solutions to (\ref{traj}) together with 68\% probability contours for Gaussian
approximations to $p(n_0,n_1)$, starting from initial delta functions.
Fig.~1(a) shows that for a slow quench, diffusion is in fact very strong; this
does not necessarily mean that the metastable state is not reached, but that it
may be reached by diffusive nucleation rather than via the critical slowing down
mechanism we are considering.  Even here, though, there are `channels' near the
axes in which diffusion is weaker.  For fast quenches, as shown in
Fig.~1(b), diffusion is clearly a small correction to predominantly systematic
evolution.  In such cases, therefore, we may obtain accurate estimates of the 
probability of reaching the metastable state, by using the linear analysis 
described above to compute the distribution $p(n_0,n_1)$ 
at some `coherent start time' $t_s\simeq\hat{t}$, and then letting the 
distribution flow under (\ref{traj}).

\begin{figure}
\epsfig{file=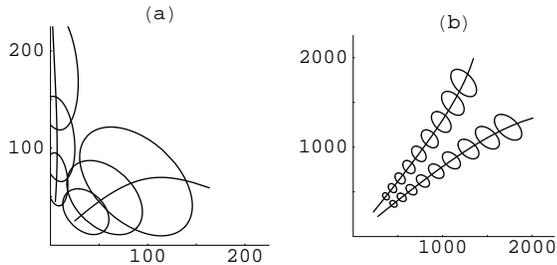, width=.9\linewidth} 
\caption{Some solutions to (\ref{traj}), with diffusion illustrated by 68\%
probability contours at selected times.  Axes are numbers of particles in modes
0 (horizontal) and 1 (vertical); starting points are on line $n_0+n_1=(2\beta
E)^{-1}$, with $\bar{n}_0(t_s)=(2\beta E)^{-1}$ setting the start time $t_s$.
Quench is $\beta\mu = \tanh(t/\tau_Q)$.  $N_c=10$; other parameters are (a)
$\beta E = 10^{-2}$, $\Gamma\tau_Q=40$, and (b) $\beta E = 10^{-3}$,
$\Gamma\tau_Q=10$.}
\label{f1}
\end{figure}

Having established that the systematic evolution of (\ref{traj}) provides a good
description, after $\hat{t}$, of a fast quench in the toy model, we can now
compare it to the TDGL evolution.  When $n_0+2n_1$ and $N_c+n_1+2n_0$ 
are both close to $N_c\mu /E$, or for low enough particle
numbers, the first line in each equation of (\ref{traj}) is indeed equivalent to
a TDGL equation (as may be seen by replacing
$n_j\to|\psi_j|^2$).  But the second line in each equation is not of
Ginzburg-Landau form:  it does not involve $\mu$, and the expression it implies
for $\dot{\psi}_j$ is not a gradient with respect to $\psi_j^*$.  These non-GL
terms conserve $n_0+n_1$, and describe doubly Bose-enhanced dissipation due to
scattering of reservoir particles off the condensate.  They turn out to imply
that the system equilibrates in energy faster than it equilibrates in particle
number.

Some representative solutions to (\ref{traj}) are shown in Fig.~2, together with
the $|\psi_j|^2$ given by the TDGL equation.  It is clear that for sufficiently
fast quenches, the two theories accord quite well, but that for slower quenches
TDGL significantly overestimates the probability of reaching the metastable
state.  If our two modes are taken to be different Fourier modes in a toroidal
trap, the vorticity of a state is simply the vorticity of the more populated
mode, so that the line $n_0=n_1$ is the border between vorticities; all initial
points in Fig.~2 are above this line.  So not even TDGL evolution conserves
vorticity, but the QKT evolution changes vorticity more easily, especially for
slower quenches.

\begin{figure}
 \epsfig{file=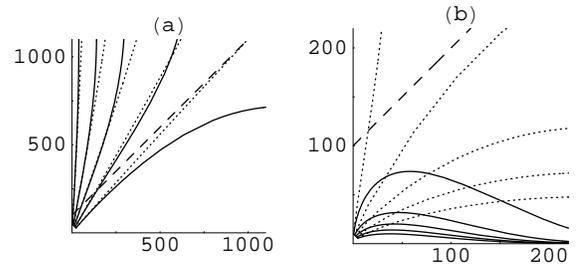, width=.9\linewidth} 
 \caption{Trajectories from QKT (solid) and TDGL (dotted); heavy dashed line is 
threshold for metastability of 
mode 1.  Initial times are $\hat{t}$; quench is $\beta\mu=\tanh(t/\tau_Q)$, 
$\beta=\beta_c e^{\tanh(t/\tau_Q)}$.  
Parameters are $N_c=100$, and (a) $\Gamma\tau_Q=10$, $\beta_c E=0.01$; (b) 
$\Gamma\tau_Q = 100$, $\beta_c E = 0.05$.}
 \label{f2}
 \end{figure}

Despite the shortcomings of TDGL revealed by our toy model, we would
like to emphasize that in fact QKT does show that
TDGL is relevant to trapped dilute gases, even very far from equilibrium:  
what TDGL
requires is not outright rejection, but corrections, from diffusion and
dissipation.  And although these corrections may be substantial, the gross
features predicted by TDGL are still recovered, with faster quenches
and smaller biases.  While the extension of quantum kinetic theory
beyond toy models, to realistic
descriptions of topological defect formation, will obviously require
much further study, we believe that the prospects for experimental realization
of spontaneous defects, as predicted by the Ginzburg-Landau theory, are very
encouraging.

This research was supported in part by the National Science Foundation under 
Grant No. PHY94-07194.


\begin{references}

\bibitem{BECexp} M. Anderson {\it et al.}, Science {\bf 269}, 198 (1995); 
K.B.~Davis {\it et al.}, Phys. Rev. Lett. {\bf 75}, 3969 (1995); C.C.~Bradley 
{\it et al.}, Phys. Rev. Lett. {\bf 75}, 1687 (1995).  

\bibitem{WHZ} W.H.~Zurek, Nature {\bf 317}, 505 (1985); Acta Physica Polonica 
{\bf B 24}, 1301 (1993); Phys. Rep. {\bf 276}, 177 (1996).

\bibitem{Kibble} T.W.B.~Kibble, J. Phys.  {\bf A 9}, 1387 (1976).

\bibitem{footnote} The apparent violation of angular momentum 
is resolved by noting that the superfluid is not isolated.

\bibitem{HE3exp} P.C.~Hendry {\it et al.}, Nature {\bf 368}, 315 (1994); 
V.M.H.~Ruutu {\it et al.}, Nature {\bf 382}, 332 (1996); C.~Ba\"uerle {\it et 
al.}, Nature {\bf 382}, 334 (1996).

\bibitem{numerical} P.~Laguna and W.H.~Zurek, Phys. Rev. Lett. {\bf 78}, 
2519 (1997); hep-ph/9711411; A. Yates and W.H. Zurek, hep-ph/9801223, to appear 
in Physical Review Letters.

\bibitem{QKT} C.W.~Gardiner and P.~Zoller, Phys. Rev. {\bf A55}, 2901 (1997); 
cond-mat/9712002; D.~Jaksch, C.W.~Gardiner, and P. Zoller, Phys. Rev. {\bf A56}, 
575 (1997); J.R.~Anglin, Phys. Rev. Lett. {\bf 79}, 6 (1997). 


\end{references}
\end{document}